\documentclass[conference]{IEEEtran}
\usepackage[caption=false]{subfig}
\usepackage{cite}
\usepackage{amsmath,amssymb,amsfonts}
\usepackage{algorithm}
\usepackage{algpseudocode}
\usepackage{graphicx}
\usepackage{textcomp}
\usepackage{xcolor}
\usepackage{multirow}
\usepackage{hyperref}
\usepackage{listings}
\usepackage{booktabs}
\usepackage{array,tabularx,makecell}
\newcolumntype{C}[1]{>{\centering\arraybackslash}p{#1}} % centered fixed-width col
\newcolumntype{R}[1]{>{\raggedleft\arraybackslash}p{#1}} % right-aligned fixed-width col

\IEEEoverridecommandlockouts

\def\BibTeX{{\rm B\kern-.05em{\sc i\kern-.025em b}\kern-.08em
    T\kern-.1667em\lower.7ex\hbox{E}\kern-.125emX}}
\begin{document}

\lstdefinestyle{c-style}{
    language=C,
    basicstyle=\ttfamily\small,
    keywordstyle=\color{blue},
    commentstyle=\color{gray},
    stringstyle=\color{orange},
    numbers=left,
    numberstyle=\tiny\color{gray},
    stepnumber=1,
    numbersep=10pt,
    backgroundcolor=\color{white},
    showspaces=false,
    showstringspaces=false,
    showtabs=false,
    frame=single,
    tabsize=4,
    captionpos=b,
    breaklines=true,
    breakatwhitespace=false,
    escapeinside={(*@}{@*)}
}

\lstdefinestyle{verilog-style}{
  language=Verilog,
  basicstyle=\ttfamily\small,
  keywordstyle=\color{blue},
  commentstyle=\color{gray},
  numbers=left,
  numberstyle=\tiny\color{gray},
  stepnumber=1,
  numbersep=10pt,
  backgroundcolor=\color{white},
  showspaces=false,
  showstringspaces=false,
  showtabs=false,
  frame=single,
  tabsize=4,
  captionpos=b,
  breaklines=true,
  breakatwhitespace=false
}

\renewcommand{\algorithmicrequire}{\textbf{Input:}}
\renewcommand{\algorithmicensure}{\textbf{Output:}}

\newcommand{\redtext}[1]{\textcolor{red}{#1}}

\newif\ifcomment
\commenttrue

\newcommand{\yf}[1]{\ifcomment{\color{blue}
\emph{[FEI: #1]}}\fi}
\newcommand{\ding}[1]{\ifcomment{\color{green}
\emph{[AD: #1]}}\fi}
\newcommand{\davis}[1]{\ifcomment{\color{red}
\emph{[DR: #1]}}\fi}
\newcommand{\rev}[1]{\ifcomment{\color{blue}#1}\fi}

\title{Exploring Side-Channel Protections in Hardware Implementations of PQC ML-KEM Verification}

\author{
\IEEEauthorblockN{Davis Ranney\IEEEauthorrefmark{1}, Yashaswini I Makaram\IEEEauthorrefmark{1}, A. Adam Ding\IEEEauthorrefmark{2}, Yunsi Fei\IEEEauthorrefmark{1}}
\IEEEauthorblockA{\IEEEauthorrefmark{1}Electrical and Computer Engineering, Northeastern University, Boston, USA \\
\IEEEauthorrefmark{2}Mathematics, Northeastern University, Boston, USA \\
\{ranney.d, imakaram.y, a.ding, y.fei\}@northeastern.edu}
\thanks{This work was supported in part by the National Science Foundation (NSF) under grants SaTC-1929300, CNS-1916762, and industry partners of NSF IUCRC CHEST.}
}

\maketitle
\begin{center}
\large\textbf{Practical Experience Report}
\end{center}

\begin{abstract}
As ML-KEM is adopted as a post-quantum cryptographic standard, resilience against physical side-channel attacks has become essential. 
Among the constituent steps, the decapsulation Fujisaki-Okamoto (FO) verification is particularly vulnerable to side-channel power and electromagnetic (EM) analysis. 
In this work, we focus on common FPGA-based implementations and examine their side-channel vulnerabilities, and compare them with those of microcontroller implementations.
Three verification implementations, unprotected, hash-based (first-order), and higher-order masked, are evaluated for side-channel security on both a microcontroller and an FPGA. 
While FPGAs offer higher speed and parallelism, they often exhibit stronger side-channel leakage, especially in high bandwidth configurations. 
The higher-order masked designs still leak information about the underlying data due to hardware-level effects and data-dependent processing. 
Our experiments show that their parallelized processing on FPGAs introduces sufficient first-order leakage for full secret-key recovery. 
These results underscore the persistent challenge of securing PQC algorithms in performance-constrained and parallelized hardware environments. 
\end{abstract}
\section{Introduction}
\label{sec:introduction}
The newly established Post-Quantum Cryptography (PQC) standards, such as the Module-Lattice-based Key-Encapsulation Mechanism (ML-KEM, based on CRYSTALS-Kyber algorithm) ~\cite{national_institute_of_standards_and_technology_us_module-lattice-based_2024}, are critical for securing communication against quantum-capable adversaries.
ML-KEM enables two parties to derive a symmetric key and engage in encrypted communication \cite{bos_crystals_2018}, leveraging the hardness of lattice problems and Module Learning with Errors to resist against quantum computers. 
With the broad adoption of PQC standards in numerous systems, their implementations range from embedded devices to large servers. 
However, the current implementations of ML-KEM, particularly the Fujisaki-Okamoto (FO) verification step during decapsulation, remain vulnerable to side-channel analysis (SCA) and fault injection (FI) attacks \cite{bhasin_attacking_2021, danvers_higher-order_2022, danvers_revisiting_2023, hermelink_insecurity_nodate, hermelink_fault-enabled_2021, hermelink_belief_2023}. 

Specifically, the FO verification involves comparing a locally re-encrypted ciphertext with the original input. 
A detected mismatch will cause the algorithm to output random bytes, ensuring IND-CCA2 (Indistinguishability under Adaptive Chosen-Ciphertext Attack) security. 
However, prior research has identified that this verification step can serve as an oracle, enabling attackers to recover the private key used in the decryption by inputting chosen ciphertexts, which are specifically crafted and only differ slightly from honestly generated ones \cite{danvers_higher-order_2022, hermelink_fault-enabled_2021, hermelink_belief_2023}. 
These attacks utilize electromagnetic (EM) emanations or power consumption during the comparison step to distinguish between the cases where the re-generated ciphertext differs minimally or significantly from the original input, yielding two different inequalities that involve the secret key. 
By systematically classifying these ciphertext variations, adversaries can build a structured set of inequalities that can be solved through belief propagation to recover an entire long-term secret key. 
Consequently, any session key established by the ML-KEM will be recovered, breaching the confidentiality of the messages and communication encrypted by the session key \cite{hermelink_fault-enabled_2021, hermelink_belief_2023}.
This also enables adversaries to impersonate the legitimate communicating parties.

\textbf{Contribution:} Previous research primarily demonstrated these vulnerabilities on software and microcontroller platforms \cite{danvers_higher-order_2022, danvers_revisiting_2023, hermelink_insecurity_nodate}, and proposed masking-based countermeasures to mitigate side-channel EM or power leakage. 
In this work, we examine whether FPGA-based implementations of ML-KEM can mitigate such vulnerabilities. 
We implement ML-KEM verification, both unprotected and protected with different strengths, on an FPGA and systematically evaluate its susceptibility to power and electromagnetic side-channel attacks. 
Our analysis shows that compared to microcontrollers, FPGA implementations do not inherently reduce these risks and, due to parallelism introduced for performance, even yield stronger and more distinguishable leakage signals.
Our results underscore the necessity of more robust protections against side-channel attacks for ML-KEM.
\section{Background}
\label{sec:background}

\subsection{ML-KEM}
\label{sec:mlkem}

ML-KEM, based on CRYSTALS-Kyber, is a PQC algorithm standardized by NIST \cite{computer_security_division_announcing_2022}. 
ML-KEM comprises three phases: key generation, encapsulation, and decapsulation \cite{national_institute_of_standards_and_technology_us_module-lattice-based_2024}. 
During key generation, a public-private key pair is created using securely generated random values. 
Encapsulation then employs the public key to encrypt a ciphertext based on a message, which is securely sent to the recipient. 
It also locally generates a shared secret based on the message and the public key. 
Decapsulation utilizes the private key to decrypt the input ciphertext and generate the shared secret, enabling secure communication between the two parties. 

ML-KEM's decapsulation integrates the Fujisaki-Okamoto (FO) transform to ensure resistance to adaptive chosen-ciphertext attacks to be IND-CCA2 secure. 
The FO transform re-encrypts the message decrypted from the ciphertext to generate a recovered ciphertext and compares it with the input ciphertext. 
If they match, it returns a valid shared secret; otherwise, it generates a random output. 
Algorithm \ref{alg:mlkem-decapsulate} shows the algorithm's design abstractly.
Although strengthening the security of ML-KEM against traditional algorithmic cryptoanalysis, this verification step presents vulnerabilities that can be exploited through physical side-channel attacks.

\begin{algorithm}[ht]
\caption{ML-KEM Decapsulate}
\label{alg:mlkem-decapsulate}
\begin{algorithmic}[1]
\Require Public Key $pk$, Private Key $sk$, Ciphertext $c$
\Ensure Shared Secret $K$, 
\State $h \gets \texttt{Hash-H}(pk)$
\State $z \gets \texttt{SampleRejectionKey}()$
\State $m' \gets \texttt{Decrypt}(sk, c)$
\State $K' \gets \texttt{Hash-G}(m', h)$
\State $\overline{K} \gets \texttt{Hash-J}(z, c)$ // z is a random seed
\State $c' \gets \texttt{Encrypt}(pk, m')$
\State \textcolor{red}{\textbf{if} $c \neq c'$ \textbf{then}}
\State \hspace{1em}$K' \gets \overline{K}$ // copy implicit rejection key
\State \textbf{end if}
\State \textbf{return} $K'$
\end{algorithmic}
\end{algorithm}

\subsection{Previous Side Channel Attacks on FO Verification}
\label{sec:related}
Physical side-channel leakage of the FO transform can be exploited to recover the private key, even when it is coded to ensure constant-time execution \cite{danvers_timing_2019, ravi_generic_2020}. 
In these side-channel analysis attacks, an adversary crafts slightly malformed ciphertexts from honestly generated ciphertexts, such that during verification, the recomputed ciphertext is either only slightly different or significantly different from the original \cite{oder_practical_2018}.
The former case is regarded as a decapsulation \textbf{success} and the latter one a decapsulation \textbf{failure}.
Being able to classify these different input ciphertexts allows attackers to recover information about the private key \cite{hermelink_fault-enabled_2021}.
EM or power side-channel analysis can distinguish between these two classes because at certain time points the side-channel leakage value differs significantly between the success and failure cases, with small versus large values, respectively \cite{danvers_higher-order_2022}. 
By collecting many traces and ciphertexts, an attacker can construct a system of inequalities that can be solved to recover the private key definitively \cite{hermelink_belief_2023}.
Under this attack model, the attacker needs physical access to the device to conduct measurements and to send an arbitrary number of ciphertexts to the device using the same secret key.

Initial approaches to protect the verification step from side-channel analysis sought to obscure the ciphertexts under comparison by computing their cryptographic hashes and comparing the resulting hash values rather than performing a direct comparison \cite{oder_practical_2018}. 
However, it has been demonstrated that such hash-based comparisons remain vulnerable to horizontal differential power attacks \cite{bhasin_attacking_2021}. 
These attacks exploit the deterministic and repeatable structure of hash function execution to correlate with specific power or EM signatures. 
Specifically, SHAKE-128 (SHA-3) \cite{national_institute_of_standards_and_technology_us_sha-3_2015} performs multiple rounds of permutation on the same state block.
The early rounds retain structural similarities to the original input, and the differences gradually propagate and amplify throughout the rounds. 
By computing a point-wise differential trace between the two executions of the hash function on the two ciphertexts in the first hashing round, attackers can effectively perform a simple power analysis (SPA) attack against this first-order-protected verification, distinguishing between ciphertexts that differ slightly or significantly.
Previous attacks against this type of hashed comparison can be described as side-channel collision attacks \cite{danvers_higher-order_2022}. 

In response, more sophisticated protections were developed, including high-order masking schemes designed to split ciphertexts and intermediate values into multiple randomized shares \cite{bhasin_attacking_2021, danvers_revisiting_2023}. 
These techniques distribute sensitive data across several computation branches with substantial randomization, theoretically obscuring any single point of leakage. 
However, practical implementations and optimizations have introduced new vulnerabilities \cite{hermelink_insecurity_nodate}.
For example, performance optimizations to reduce some of the masking overhead, as proposed in \cite{danvers_revisiting_2023}, rely on reduction operations whose behavior depends on the presence of one-bits in the processed values. 
Even though the implementation is secure against t-probing, the leakage from each random share is sufficiently pronounced that an attacker can still distinguish between decapsulation success and failure with high probability \cite{hermelink_insecurity_nodate}. 

Previous attacks in this area have primarily focused on implementations running on Cortex M4-based microcontrollers, the baseline embedded platform established by NIST for evaluating PQC schemes \cite{national_institute_of_standards_and_technology_us_module-lattice-based_2024}. 
The space of hardware-accelerated, FPGA-based FO verification implementations has been largely unexplored.
For example, several other works that explore hardware acceleration of ML-KEM \cite{jati_configurable_2024, zhao_side_2023, xu_hardware-friendly_2024, fritzmann_masked_2021} focus on optimizing and protecting operations outside of the FO verification, including the number-theoretic transform (NTT), inverse NTT, modular reduction, and point-wise multiplication (PWM).
Our work is among the first to systematically investigate the side-channel security of the FO verification step and its acceleration. 
Ultimately, we revealed that parallelization yields stronger power-leakage signals and that directly parallelizing high-order masking is less effective for securing the verification step on an FPGA than on a microcontroller, indicating that bespoke solutions for FPGAs will be necessary.
\section{Experimental Approach}
\label{sec:approach}

\subsection{Motivations}
\label{sec:motivations}
Many masking methods intended to protect the FO transform's verification step from side-channel leakage significantly increase the computational complexity of this component. 
What was originally a straightforward check in unprotected implementations has become a bottleneck under masking, requiring multiple shares, modular reductions, and complex recombination logic \cite{danvers_higher-order_2022, danvers_revisiting_2023, hermelink_insecurity_nodate}.
Despite these costly additions, side-channel vulnerabilities persist.

To better understand the trade-off between implementation costs and side-channel security, we sought to explore how masking schemes and FPGA implementations affect both the side-channel leakage and noise profiles, while improving the performance of protected implementations.
Our goal was to evaluate whether these hardware-based approaches provide measurable advantages or disadvantages compared to software implementations, particularly in terms of signal-to-noise ratio (SNR) and overall performance.

Specifically, we evaluated how several important algorithm and hardware factors, including pipelining and parallel processing elements, the number of shares used in masking, and the randomization of independent sub-functions, affect an attacker's ability to distinguish between two kinds of input ciphertext. 
This methodology provides insight into how masking strategies interact with hardware-level behavior and identifies the conditions under which FPGA-based deployments may offer either security benefits or remain vulnerable to side-channel attacks.

To evaluate side-channel leakage of both software and hardware implementations, we conducted experiments on two platforms. 
First, we used the RISCURE \textit{Pinata} board, which features an ARM Cortex-M4 microcontroller operating at 168 MHz \cite{keysight_ds1030a_nodate}. 
This platform represents the embedded profile recommended by NIST for evaluating PQC schemes and serves as a point of comparison with prior side-channel research.

For hardware-based evaluation, we utilized a \textit{SAKURA-G} FPGA board, which is equipped with an Xilinx Spartan-6 chip and includes built-in amplification circuitry specifically designed for \textbf{power side-channel analysis} \cite{noauthor_sakura_nodate}. 
This platform enabled us to evaluate how parallelism, routing, and synthesis-level optimizations impact side-channel leakage in more spatially distributed designs.

Traces were captured using a Teledyne LeCroy WaveRunner 6 Zi oscilloscope configured to sample at 1 GS/s \cite{noauthor_buy_nodate}. 
Each board was connected to a host computer that coordinated data transmission and collection. 
The computer generated all required randomness and test vectors, which were sent to the device under test to ensure that each implementation performed only the relevant function under side-channel observation. 
This setup minimized extraneous computation and isolated the comparison and reduction logic to capture leakage with maximum fidelity.

\subsection{Unprotected Implementation}
\label{sec:unprotected}

Our first implementation served as the unprotected baseline for both platforms. 
In the case of ML-KEM-512, the ciphertext to be compared during verification is 768 bytes in length \cite{national_institute_of_standards_and_technology_us_module-lattice-based_2024}. 
For this type of attack, ML-KEM-512, ML-KEM-768, and ML-KEM-1024 differ only in ciphertext length and do not change the underlying verification computations.
Therefore, we decided to evaluate our attacks only using ML-KEM-512, as it provides the minimum security thresholds for ML-KEM. 
On the Cortex-M4 microcontroller, the comparison between the original and recomputed ciphertexts is implemented as a simple loop using byte-wise XOR and AND operations. 
Each byte of the two ciphertexts is processed sequentially across 768 iterations.
On the FPGA, we designed a flexible comparison module that compares ciphertext bytes in parallel at varying levels of granularity. 
This module supports comparisons at widths ranging from 1 byte up to 64 bytes per clock cycle. 
This flexibility enabled us to investigate the effect of varying levels of parallelization on the resulting side-channel leakage. 
These unprotected implementations served as a baseline for measuring the impact of the inherent noise and execution characteristics of microcontroller and FPGA platforms on leakage observability. 

\subsection{Hash-Based Comparison Implementation}
\label{sec:hashedcompare}
We implemented a first-order countermeasure for FO verification, hashing the two ciphertexts with SHAKE-128 (SHA-3) \cite{national_institute_of_standards_and_technology_us_sha-3_2015} before comparison, as originally proposed in \cite{oder_practical_2018}. 
Each 768-byte ciphertext was hashed individually to produce a fixed 128-byte digest. 
The resulting hashes were then compared using similar XOR and OR-based logic as in the unprotected implementation.
The hashing algorithm propagated minor bit differences so that any difference would result in a significantly different hash value.
On the FPGA, we customized a SHAKE-128 implementation tailored for the fixed 768-byte inputs produced by ML-KEM-512. 
SHAKE128 has an input size limit of 168 bytes per iteration, and new data can be 'absorbed' into the state over multiple iterations.
Our design executed the hash core across five absorb cycles to handle the full input, followed by a squeezing step at the end to produce the 128-bit output. 
Due to synthesis/resource constraints on the Spartan-6 FPGA, our implementation could not process the entire 1600-bit Keccak state (organized in three dimensions, 5$\times$5$\times$64 for row$\times$column$\times$lane) in parallel, which would have enabled the highest-throughput design. 
Instead, the design was pipelined at the plane level, the next-highest level of parallelization, in which each 320-bit plane of the state matrix was updated at once. 
% \rev{(Shortened the previous paragraph and moved the information to the previous section)}

Another approach to mitigate side-channel threats is to randomize the execution order of operations, thereby introducing temporal entropy to hide the leakage. 
In our FPGA implementation, we experimented with randomizing the row-processing order within each round while ensuring the output remained functionally equivalent. 
This form of lightweight shuffling was designed to obscure consistent temporal patterns in power and EM traces while preserving compatibility with the core Keccak structure. 
However, as shown in the next section, our evaluation demonstrates that such hash-based comparison with random row shuffling fails to provide adequate protection against side-channel analysis in this setting.

\subsection{Higher Order Masked Implementation}
\label{sec:higherorder}
To explore higher-order protection techniques, we implemented a masked verification algorithm based on the Galois Field (GF)-optimized approach previously designed by \cite{danvers_revisiting_2023}. 
This design serves as our reference software implementation on the Cortex-M4 microcontroller. 
For the FPGA, we developed a pipeline-optimized version of this higher-order comparison. 
Due to hardware complexity constraints, we focused specifically on the portion of the verification step identified as leaky in prior work by \cite{hermelink_insecurity_nodate}, namely, the masked reduction and comparison stage. 
Our FPGA implementation applied the GF-based operations in a pipelined stage, parallelizing across the configurable number of shares.
The design was synthesized to distribute the logic spatially, thereby minimizing leakage. 
This allowed us to evaluate the inherent difficulty of securely executing higher-order masked comparisons on hardware platforms, particularly under realistic timing and resource constraints.
\section{Experimental Results}
We present evaluation results for the three implementations, covering side-channel leakage and performance. 

\subsection{Unprotected Verification and Noise Profiling}
\label{sec:noise}
Before performing any side-channel analysis, we preprocessed all collected traces to ensure high fidelity and consistency. 
This included temporal alignment of the traces to account for clock drift or communication jitter, ensuring that corresponding processing cycles across traces were aligned correctly. 
We then filtered out any traces that were corrupted, incomplete, or produced no usable signal, typically due to intermittent transmission issues, noise bursts, or invalid ciphertext input conditions. 

\begin{figure}[b]
    \centering
    \includegraphics[width=0.8\linewidth]{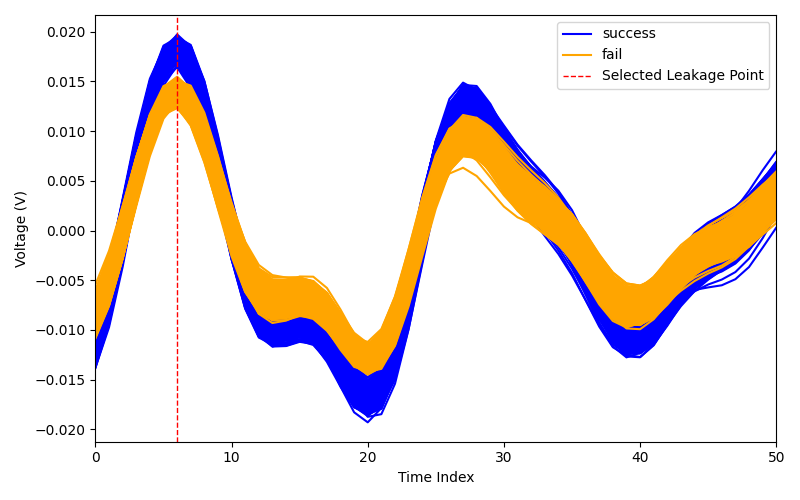}
    \caption{Processed traces for unprotected FPGA 128-bit (comparison width) implementation}
    \label{fig:filtered_trace}
\end{figure}

We collected and processed 8,000 power traces of each class, and the traces are visualized in \autoref{fig:filtered_trace}, where a time point is selected with the best distinguishability/classification. 
We used T-testing to determine which time point in the power trace yielded the best leakage.
We generated a set of inequalities based on the classification results and solved for the private key using the belief-propagation solver described in \cite{hermelink_belief_2023}.
That belief propagation solver can solve for an entire ML-KEM-512 key using 6,000 inequalities with perfectly classified traces and up to 8,000 inequalities when the success rate drops to 90\%.

We calculated the signal-to-noise ratio (SNR) of the power traces to evaluate the quality of the captured side-channel signals. 
To quantify leakage at our selected time index \(t\), we computed a signal-to-noise ratio (SNR) for classifying two classes.
For each class (success/failure), we extracted the data at the selected time index \(t\) into two vectors and computed the mean and unbiased variance. 
The “signal” was defined as the absolute difference between class means, and the “noise” as the square root of the sum of class variances, yielding
\begin{equation}
\mathrm{SNR}(t)=
\frac{\bigl|\mu_{\mathrm{s}}(t)-\mu_{\mathrm{f}}(t)\bigr|}
{\sqrt{\sigma_{\mathrm{s}}^2(t)+\sigma_{\mathrm{f}}^2(t)}}.
\label{eq:snr}
\end{equation}
where higher SNR values indicate clearer, more exploitable leakage.

\begin{figure*}[ht]
    \centering
    \subfloat[Microcontroller (SNR=2.40)]{\includegraphics[width=0.4\textwidth]{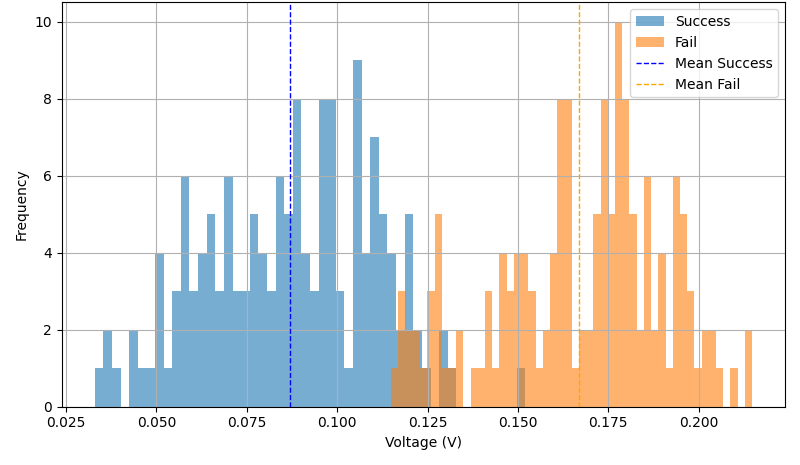}}
    \subfloat[FPGA 32-Bit (SNR=0.95)]{\includegraphics[width=0.4\textwidth]{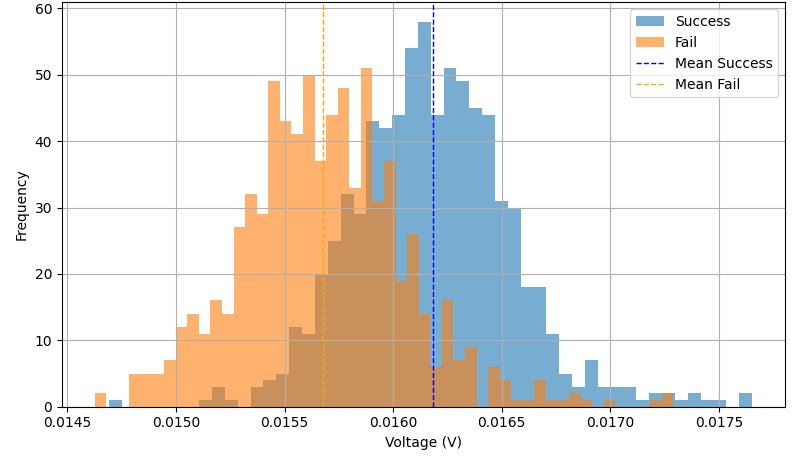}}
    \hfill
    \subfloat[FPGA 128-Bit (SNR=3.83)]{\includegraphics[width=0.4\textwidth]{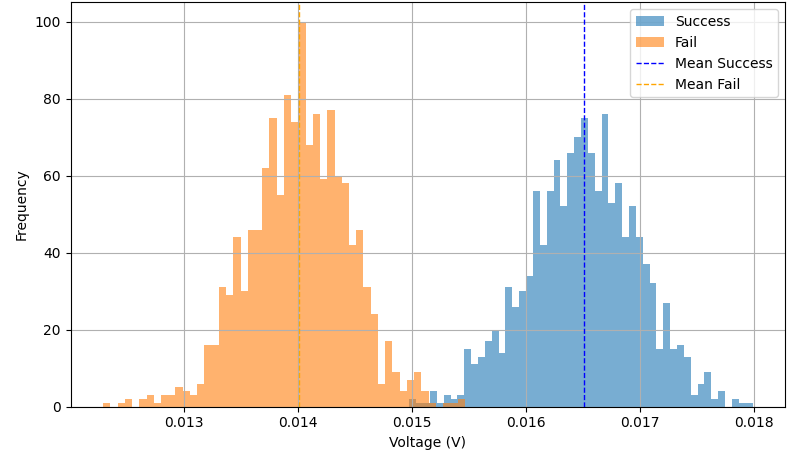}}
    \subfloat[FPGA 512-Bit (SNR=8.33)]{\includegraphics[width=0.4\textwidth]{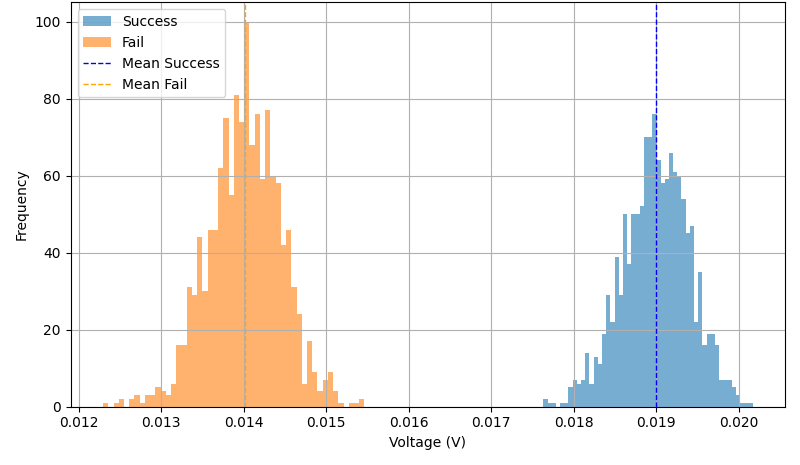}}
    \hfill
    \caption{Power distributions of the two classes and SNR analysis of unprotected verification implementations on microcontroller and FPGA}
    \label{fig:snr_compared}
\end{figure*}

We compared SNR across multiple implementations, including a microcontroller and an FPGA implementation with varying degrees of parallelization for 32-bit, 128-bit, and 512-bit data widths. 
The SNRs and the two-class distributions are presented in \autoref{fig:snr_compared}.
The SNR of the microcontroller implementation is $2.40$. Among the FPGA implementations, only the minimal-parallelization (32-bit comparison width) implementation has an SNR of $0.95$, while the others have higher SNRs.

With an appropriate threshold applied, when classifying the ciphertexts into decapsulation success and failure classes based on the power values, we achieved a \textbf{classification accuracy} of \textbf{94.8\%} for the microcontroller, as shown in the last column of Table~\ref{tab:fpga_utilization_unprotected}. 
The classification accuracy drops to \textbf{62.9\%} for the 32-bit FPGA implementation. 
With the belief propagation solver \cite{hermelink_belief_2023}, this is insufficient for the solver to recover the entire secret key successfully. 
With our empirical results, we consider any side-channel attack with a classification accuracy below \textbf{80\%} to be unsuccessful \cite{hermelink_insecurity_nodate}.

\begin{table}[t]
\centering
\caption{Cross-platform Comparison of Resource Utilization, Throughput, and Side-channel Classifier Accuracy}
\label{tab:fpga_utilization_unprotected}
\setlength{\tabcolsep}{3pt} % tighten column padding
\begin{tabularx}{\linewidth}{l|R{8mm}|R{8mm}|R{8mm}|R{16mm}|R{17mm}}
\toprule
\textbf{Bit-Width} & \textbf{LUTs} & \textbf{FFs} & \textbf{Slices} &
\makecell{\textbf{Throughput}\\\textbf{(Gb/s)}} &
\makecell{\textbf{Classification}\\\textbf{Accuracy} (\%)} \\
\midrule
32-bit  FPGA    &  211  & 236  &  82  &  1.536 &  62.9  \\
128-bit FPGA    &  319  & 349  & 110  &  6.144 &  99.1  \\
512-bit FPGA    & 1262  & 1288 & 390  & 24.576 & 100.0  \\
Microcontroller &   --  &  --  &  --  &  0.046 &  94.8  \\
\bottomrule
\end{tabularx}
\end{table}

When the width of parallelism increased to 128-bit and 512-bit, the SNR increased significantly to $3.83$ and $8.33$, respectively, much higher than those of the microcontroller implementation, as shown in \autoref{fig:snr_compared}, demonstrating that the wider combinational logic in the FPGA correlates with clearer and more distinguishable side-channel signals. 
With the larger bit-widths, the classification accuracy increased to \textbf{99.1\%} and \textbf{100.0\%}, respectively, for the 128-bit and 512-bit FPGA implementations, as shown in the last column of Table~\ref{tab:fpga_utilization_unprotected}. 
These results can be explained by the significant increase in the signal and reduction of noise due to wider comparisons.
With a wider implementation, the mean power for the success class does not change much (with a Hamming weight of 0 or 1), while the mean power for the failure class increases significantly (the average Hamming weight being half of the comparison width), resulting in a larger differential power/signal. 
We also speculate that the noise level of the wider comparison implementation is reduced due to the averaging effect. 
Consequently, our SNR and classification accuracy improve proportionally to the FPGA bit-width. 

We also compared resource utilization and throughput across implementations, and the results are presented in \autoref{tab:fpga_utilization_unprotected}. 
The FPGA achieved a substantial performance improvement over the microcontroller, with throughput scaling roughly linearly with processing width. 
At the widest tested configuration (512-bit), the FPGA achieved a speedup of approximately \textbf{530x} over the microcontroller's baseline throughput.
However, there is a trade-off: as FPGA parallelism increases, side-channel observability also grows, making higher-performance configurations more vulnerable to side-channel leakage.

\subsection{Hash-Based Implementation}
\label{sec:hashresults}

For the hashed comparison implementation using SHAKE-128, the FPGA executed each of the five Keccak permutation steps using row-by-row parallelization. 
This design allowed us to process 320 bits of ciphertext at a time.
We verified that the output from our FPGA implementation matches our reference and collected 8,000 traces to solve for the key as described in Section \ref{sec:noise}.
After performing the similar alignment and filtering procedures described previously in Section \ref{sec:unprotected}, we constructed difference traces between pairs of ciphertexts being hashed and classified them to evaluate side-channel distinguishability. 

\begin{figure}[h]
    \centering
    \includegraphics[width=0.8\linewidth]{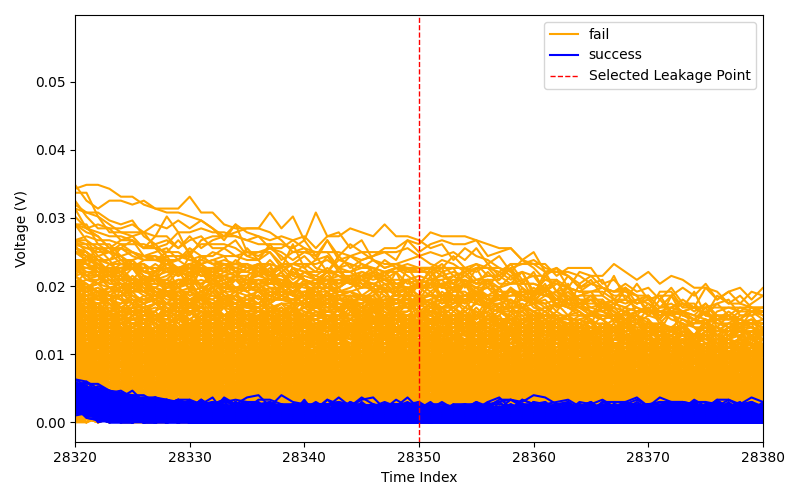}
    \caption{Distribution of absolute power differences power traces from FPGA hashed implementations}
    \label{fig:hash_distribution}
\end{figure}

To quantify this, we performed T-tests across the aligned difference traces. 
The t-test results reveal that the Rho/Pi subfunction in the Keccak core has the highest leakage within the first Keccak round. 
Zooming in on the clock cycle with the strongest statistical evidence of leakage, \autoref{fig:hash_distribution} presents the raw difference traces, with the most prominent leakage location highlighted by the vertical red dotted line. 
This point occurred during the early stages of the permutation, when the state matrix still retained structural dependence on the input data.

\begin{figure}[t]
    \centering
    \includegraphics[width=0.8\linewidth]{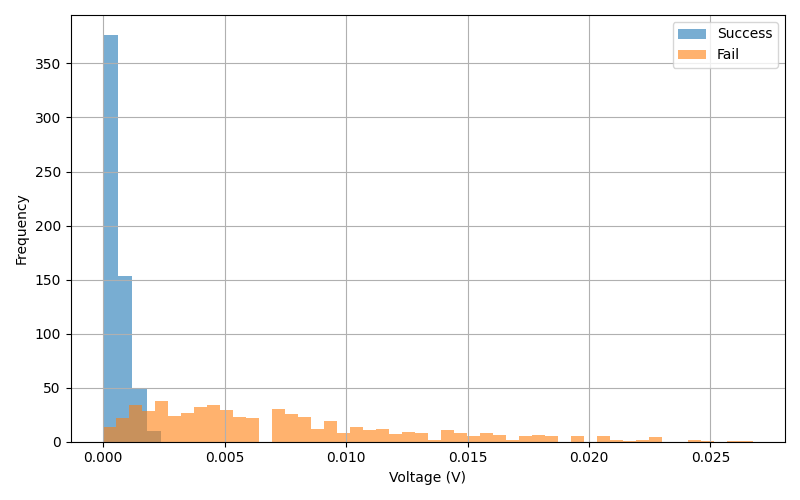}
    \caption{Distribution of absolute power differences power values from FPGA hash-based attack for the selected leakage time point}
    \label{fig:hash_hist}
\end{figure}

Examining the distribution of power traces at this clock cycle and leakage location (\autoref{fig:hash_distribution}, \autoref{fig:hash_hist}) reveals that, beyond a simple mean difference, failure cases exhibit a much broader spread of power values compared to the tightly clustered success cases. 
This separation enables straightforward classification of decapsulation success versus failure using power samples at the leakage point. 
With a clustering-based machine learning algorithm, we achieved a classification accuracy of \textbf{94.7\%}.

\begin{figure}[b]
    \centering
    \includegraphics[width=0.8\linewidth]{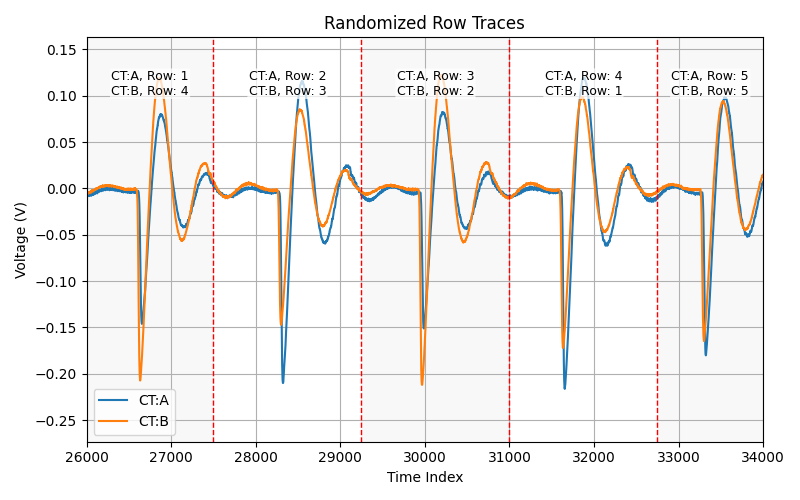}
    \caption{Zoomed-in view of power traces of FPGA hash implementation for two ciphertexts (A versus B) with randomized order of row processing.}
    \label{fig:hash_shuffle_rows}
\end{figure}

\begin{figure}[t]
    \centering
    \includegraphics[width=0.8\linewidth]{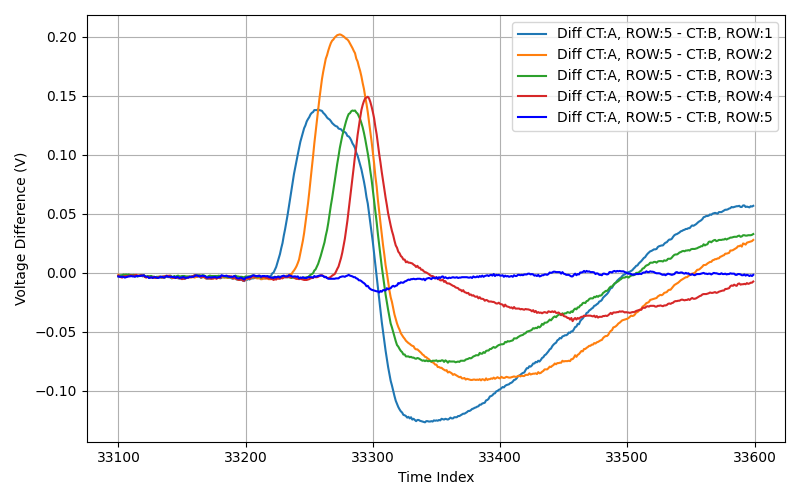}
    \caption{Difference between power traces of Ciphertext A Last cycle versus Various Cycles for Ciphertext B in FPGA Hash Implementation}
    \label{fig:hash_shuffle_diff}
\end{figure}

Furthermore, we implemented the randomized row scheduling strategy and investigated its effect on mitigating side-channel observability. 
The runtime randomization of execution order permutes power values associated with the five operations within five segments of the power traces. 
However, it is very easy to identify the permuted order by inspecting the resulting power traces.
\autoref{fig:hash_shuffle_rows} depicts two traces A and B for two ciphertexts randomized with differing execution order. 
Because the overall structure and magnitude of these five segments are very different, a simple profiling can generate a fingerprint for each segment, and the execution order can be identified. 
\autoref{fig:hash_shuffle_diff} shows the difference when comparing the last segment of A trace with each of the five segments of B traces. 
The segment of B matching the same operation in A shows significantly less deviation than the other mismatched segments. 
Thus, by selecting the segment with the minimum value of difference, we identify that B's execution order is (4, 3, 2, 1, 5) when matched to an A execution order of (1, 2, 3, 4, 5). 
This simple matching strategy identifies the correct cycle to match the rows with 99.9\% accuracy. 
Hence, the randomized row scheduling does not affect isolating the key leakage point and then applies the previous classification of decapsulation success versus failure. 
Even if the randomization were increased to process each cell of the hash state array individually, which would drastically reduce the throughput, the search space for identifying matching cell pairs would remain tractable and classifiable.

These results demonstrate that, even with lightweight randomness applied to the scheduling of permutation steps, side-channel information persists in the early rounds of SHAKE-128. 
Our data confirms the vulnerability of the hash-based comparison method to practical side-channel collision-style attacks, particularly when deployed on parallelized hardware designs.

\subsection{Higher-Order Masked Implementation}
\label{sec:higherresults}

\begin{figure}[t]
    \centering
    \includegraphics[width=0.8\linewidth]{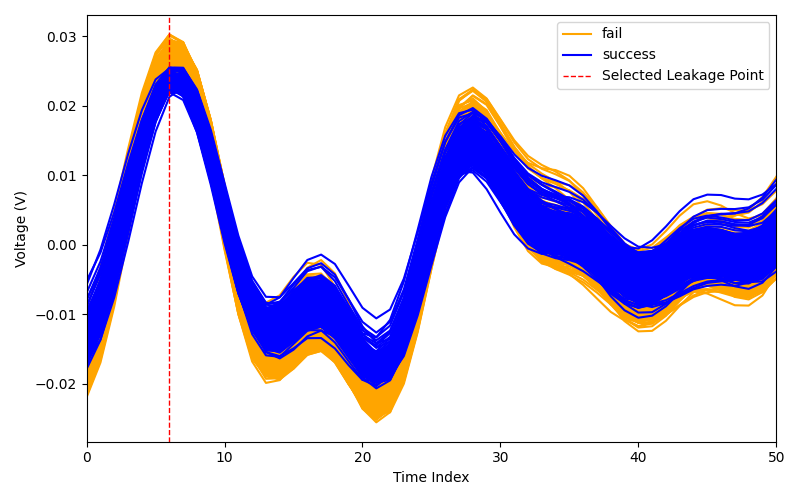}
    \caption{Filtered Power Traces for Higher Order Protected FPGA Implementation}
    \label{fig:filtered_higher_trace}
\end{figure}

The higher-order masked comparison is proven t-probing-secure, and the microcontroller implementation does not leak for orders up to t. 
However, prior work \cite{hermelink_insecurity_nodate} has shown that, even though the design utilized multiple shares and Galois Field arithmetic to obscure the sensitive data, the underlying operations remained data-dependent. 
In particular, the one-bits versus zero-bits of each share can be leaked during multiplications against fixed or semi-random constants in the GF masking structure, enabling a practical implementation of a higher-order attack.
In the highly parallelized FPGA implementation of the higher-order masked comparison, leakage from parallel multiplications effectively reduces the masking order, undermining the intended protection. 
We conducted statistical analysis on these power traces using the same methodology as for the unprotected implementation. 
\autoref{fig:filtered_higher_trace} demonstrates that decapsulation success and failure can be reliably distinguished using power values at the identified leakage location.

Using the same clustering classifier as before, we achieved a classification accuracy of \textbf{97.9\%} for the 4-share implementation, which further increased to \textbf{98.5\%} for the 6-share implementation. 
This trend arises because the higher-order reduction step processes more information per cycle as additional shares are combined, making instantaneous activity more correlated with underlying bit differences. 
Notably, both FPGA implementations achieved higher accuracy than the prior higher-order attack on microcontrollers (\textbf{95.0\%}) reported in \cite{hermelink_insecurity_nodate}. 
These results indicate that FPGA parallelization significantly undermines the side-channel resistance of higher-order masked comparison designs.
\autoref{tab:fpga_utilization_higher} summarizes the resource utilization of the FPGA and microcontroller implementations. 
The FPGA achieves a throughput that is over three orders of magnitude higher than that of the microcontroller. 
Notably, the steady-state throughput of our FPGA comparison engine remained effectively unchanged between the 4- and 6-share designs for a fixed ML-KEM parameter set. 
Since ciphertext size and overall bandwidth are constant, the additional share computations were absorbed through spatial parallelism. 
Thus, while resource utilization and per-share latency increase, the end-to-end comparison throughput remains unaffected.

\begin{table}[b]
\centering
\caption{Resource Utilization and Throughput Across Higher-Order Protected Implementations}
\label{tab:fpga_utilization_higher}
\setlength{\tabcolsep}{3pt} % tighten column padding
\begin{tabularx}{\linewidth}{l|R{7mm}|R{6mm}|R{7mm}|R{16mm}|R{17mm}}
\toprule
\textbf{\# of Shares} & \textbf{LUTs} & \textbf{FFs} & \textbf{Slices} &
\makecell{\textbf{Throughput}\\\textbf{(Gb/s)}} &
\makecell{\textbf{Classification}\\\textbf{Accuracy} (\%)} \\
\midrule
4-Share FPGA    & 14353 & 6641 & 5345 &  0.542 &  97.9   \\
6-Share FPGA    & 18565 & 8621 & 6975 &  0.542 &  98.5   \\
Microcontroller \cite{hermelink_insecurity_nodate} &  --  &  --  &  --  &  0.001 &  95.0  \\
\bottomrule
\end{tabularx}
\end{table}

In summary, the FPGA proves highly effective in accelerating masked comparisons of increasing order, which, in theory, should strengthen security. 
However, our results show the opposite effect: leakage signals grow stronger with parallelism, degrading side-channel protection even as performance improves. 
This highlights the need for far greater care when deploying higher-order masked designs on parallel hardware platforms.
\section{Alternative Countermeasures and Mitigation}
\label{sec:mitigation}

While higher-order masking remains a possible mitigation, our results show that highly parallel hardware introduces new leakage mechanisms that can weaken side-channel resistance. 
Although this work evaluates a Spartan-6 FPGA, the attack principles are not platform-specific. 
Newer FPGAs fabricated on smaller process nodes exhibit different noise characteristics but also enable greater parallelism, increasing instantaneous switching activity and amplifying leakage sources. 
As shown in our results, wider parallel implementations exhibit clearer leakage, suggesting that modern FPGA architectures may remain vulnerable despite advances in fabrication. 
Future work should evaluate how architectural differences across FPGA generations influence leakage observability.

Another proposed mitigation involves inserting dummy operations or randomized execution delays. 
While this can complicate analysis, it reduces throughput and pipeline efficiency in FPGA designs. 
More importantly, dummy operations do not eliminate leakage but only obscure it.
Given sufficient traces, attackers can filter them out using statistical classification techniques, revealing the underlying signal \cite{lee_security_2020}.

A more robust mitigation is to limit the reuse of secret keys at the system level. 
Restricting keys to short lifetimes or limited decapsulation counts reduces the number of traces available to attackers, making key recovery impractical with this attack. 
In such systems, long-term authentication can be handled by a separate mechanism, while ML-KEM keys are treated solely as ephemeral session secrets rather than as persistent root keys \cite{bos_crystals_2018}.

Finally, recent work proposes integrating masked verification earlier in the decapsulation pipeline, prior to ciphertext compression \cite{hermelink_insecurity_nodate, dubrova_breaking_nodate}. 
This enables more robust masking by distributing the verification across other decapsulation operations. 
However, these approaches remain under active investigation and require further evaluation in hardware environments.

\section{Conclusion}
\label{sec:conclusion}
This work demonstrates that even state-of-the-art countermeasures for ML-KEM’s FO verification remain susceptible to practical side-channel attacks when deployed on real hardware. 
Across a plethora of implementations, from unprotected logic to higher-order masked designs, our evaluations on both microcontroller and FPGA platforms reveal persistent and exploitable leakage.
While the performance gains and throughput of the functions were significantly better on the FPGA, our findings reveal that countermeasures developed under leakage assumptions for serial implementations of ML-KEM verification prove insufficient for parallelized FPGA implementations. 
Hardware-based protection schemes require explicit consideration of the leakage model specific to parallelized execution.

Ultimately, our results reaffirm the difficulty of implementing robust side-channel protections in hardware cryptographic modules and underscore the need to reassess what constitutes a secure-by-design implementation for post-quantum algorithms. 
As ML-KEM continues toward widespread adoption, future work must bridge the gap between theoretical security and practical deployment to ensure its long-term viability in real-world systems.

\newpage

\bibliographystyle{IEEEtran}
\bibliography{references}

\newpage

\end{document}